\documentclass[%
     11pt,
     american,
     aip,
     floatfix,
     jcp,
     reprint,
     longbibliography,
     superscriptaddress,
]{revtex4-2}

\usepackage{graphicx}
\usepackage{color}
\usepackage{amsmath}
\usepackage[final]{pdfpages}
\usepackage{tikz}
\usepackage{pgffor}

\makeatletter
\AtBeginDocument{\let\LS@rot\@undefined}
\makeatother

\makeatletter
\def\@fnsymbol#1{\ensuremath{\ifcase#1\or \dagger\or \ddagger\or
   \mathsection\or \mathparagraph\or \|\or **\or \dagger\dagger
   \or \ddagger\ddagger \else\@ctrerr\fi}}
\makeatother

\newcommand{\silica}{a-SiO$_2$}
\newcommand{\quartz}{$\alpha$-quartz}
\newcommand{\angstrom}{\text{\normalfont\AA}}

\begin{document}


\title{Graph theory-based structural analysis on density anomaly of silica glass}



\author{Aik Rui Tan}
\thanks{A.R.T. and S.U. contributed equally to this work.} 
\affiliation{Department of Materials Science and Engineering, Massachusetts Institute of Technology}

\author{Shingo Urata*}
\email{shingo.urata@agc.com}
\thanks{A.R.T. and S.U. contributed equally to this work.} 
\affiliation{Technology General Division, Planning Division, AGC Inc.}

\author{Masatsugu Yamada}
\affiliation{Innovative Technology Laboratories, AGC Inc.}

\author{Rafael G\'omez-Bombarelli*}
\email{rafagb@mit.edu}
\affiliation{Department of Materials Science and Engineering, Massachusetts Institute of Technology}



\date{\today}

\begin{abstract}
Analyzing the atomic structure of glassy materials is a tremendous challenge both experimentally and computationally, and the lack of direct, detailed insights into glass structure hinders our ability to navigate structure-property relationships. For instance, the structural origin of the density anomaly in silica glasses - the negative thermal expansion coefficient -  is still poorly understood. Simulations based on molecular dynamics (MD) produce atomically resolved structures, but quantifying the role of disorder in the density anomaly is challenging. Here, we propose to use a a graph-theoretical approach to assess topological differences between disordered structural arrangements from MD trajectories of silica glasses. A graph similarity metric quantifies the similarity between the covalent networks and can characterize the nature of the disordered solid, by comparing to reference crystalline solids, or with glasses in different thermodynamic states . This approach involves casting all-atom glass configurations as networks, and subsequently applying a graph-similarity metric (D-measure). Calculated D-measure values are then taken as the topological distances between two configurations. By measuring the topological distances of silica glass configurations across a range of temperatures, distinct structural features could be observed at temperatures higher than the fictive temperature. In addition, we compared topological distances between local atomic environments in the glass and crystalline silica phases. This approach suggests that more coesite-like and quartz-like local structures emerge in silica glasses when the density is at a minimum during the heating process. 
\end{abstract}

\maketitle

\newpage
\section{Introduction} \label{sec:introduction}

Structure-property relationships are one of the tenets of materials science\cite{Le2012,cheng_ma_2011,calas_2002,surnev_2012,pedone2018}. However, amorphous materials exhibit variable atomic topologies, which play an important role in many of their properties\cite{onodera_2020,calas_2002}. For example, a glass formed by rapid quenching will share a higher structural similarity to the high-temperature melt than a glass formed by slow cooling \cite{Debenedetti2001}. Differences in cooling rates change the ordering and configurational entropy, which in turn modifies some properties of the glass immensely. However, unlike small molecules or crystalline materials, glassy materials are structurally diverse and feature network topologies that lack translational periodicity and long-range order, which hinders precise understanding of structures that underlie their unique properties \cite{starchurski2011,rouxel2007,onodera_2020,urata2017}. In the computational domain, molecular dynamics (MD) simulations can access the abundance and diversity of local arrangements that occur in glasses, but encoding the complex topological and chemical landscape of glassy materials in a physically meaningful manner becomes a bottleneck, precluding data-driven approaches from being applied to study glassy materials efficiently\cite{chapman_2022}. 

In the case of oxide glasses, common descriptors of atomic structure include bond order of network-formers, ratio of bonding oxygen and non-bonding oxygen (BO/NBO), distributions of Q$^n$ speciation (where $n$ is number of bridging oxygens connected to a network-former cation)\cite{mysen_book,Micoulaut2006}, ring sizes, radial distribution functions and structure factors. These and others have been used to characterize and probe the structures of glasses. More recently, persistent homology has attracted attention as a novel algorithm to analyze amorphous topologies based on the vacancy distribution \cite{hiraoka2016,onodera2019,lee2017}. In terms of local structural environments, descriptors such as tetrahedral order of SiO$_4$ \cite{shell2002}, Q$^n$ speciation, and bond order for atoms\cite{Steinhardt1983} are used to represent the short-range coordination, while medium-range descriptors are still mostly limited to local strain \cite{urata2018} and potential energy \cite{takada2009, takada2010}. Analyses using these schemes to probe structural arrangements have drawn significant insights into glass properties. However, many of these methods are not able to capture small changes in the local coordination environment that may have played a distinctive role in many elusive glassy properties such as the anomalous density behavior shown in silica glass\cite{gereben1995,Tian2011}.

Silica glass has often been chosen as the representative oxide glass for study since it is in principle simple, but exhibits non-trivial behaviors. Sharing the same tetrahedral order and dynamic anomalies as water, silica exhibits a phenomenon known as density anomaly, where its density reaches a minimum at temperatures higher than the melting point, and then increases with temperature \cite{brueckner1970}. Silica also exhibits typical nonlinear behaviors such as mechanical anomaly \cite{huang2004} and fragile-to-strong transition of diffusion behavior \cite{geske2016}. These peculiar properties of silica have been studied using molecular dynamics (MD) simulations 
\cite{huang2004, geske2016, vollmayr1996, horbach1999, de2009, soules2011, jabes2012, shi2018} using a plethora of interatomic potentials such as the BKS potential\cite{van1990}, TTAM potential\cite{ttam}, and others\cite{takada2010, huang2004, de2009, soules2011, jabes2012, bertolazzo2016}. 

In this work, we propose a graph-based approach to quantify the (dis)order of atomic configurations of silica glasses obtained from MD simulations.  The approach captures local and beyond-local environments by comparing connectivity networks including second and higher order neighbors. Expressing the atomic environment as a discrete graph provides a rich signal that can easily distinguish dissimilar environments. However, quantifying the distance between two arbitrary graphs is not trivial. Here, we apply a continuous graph dissimilarity metric known as D-measure \cite{schieber2017} which ranges from 0 to 1 (isomorphic). D-measure has been applied to analyze social networks \cite{kolomeets2019,horawalavithana2019}, community structures \cite{jiang2017}, brain networks \cite{mheich2020}, and to characterize structural transition of zeolite nanoporous materials \cite{schwalbe2019}.

In Section \ref{sec:results}, we discuss the density anomalous behavior in our simulated glass systems as well as the applicability and comparison of D-measure method with another physics-informed, data-drive descriptor of atomic structure, the Smooth Overlap of Atomic Positions (SOAP),\cite{bartok2013} which has found success in a number of applications \cite{Deringer2017, De2016,Dragoni2018}. Section \ref{sec:conclusions} summarizes the findings, while, Section \ref{sec:methods} presents briefly the applied methods.

\section{Results and Discussion} \label{sec:results}

\subsection{Density anomaly} \label{subsec:density_anomaly}
The atomistic configurations analyzed in this work were obtained from classical MD simulations of silica glasses. Two different force-matching potentials were utilized to create rigorous all-atom representations of the glass structure (see Methods section for details). The first force-matching potential (named FMP-v1 hereafter) produces silica glass trajectories that exhibit a density minimum at just above 2000~K, but overestimates density of silica at room temperature by approximately 0.32 $g/cm^3$\cite{urata2021a}. On the other hand, the second force-matching potential (FMP-v2), which was developed to remedy density overestimation in FMP-v1, reproduces density of silica accurately, but is unable to reproduce the density minimum above 2000~K in the simulated silica glass trajectories\cite{urata2021b}. These two potentials are chosen specifically for study because they are optimized on the same DFT data but their application in MD produces two different systems with distinct temperature-dependence behavior. 

\begin{figure}[th]
    \begin{center}
    \includegraphics[width=\linewidth]{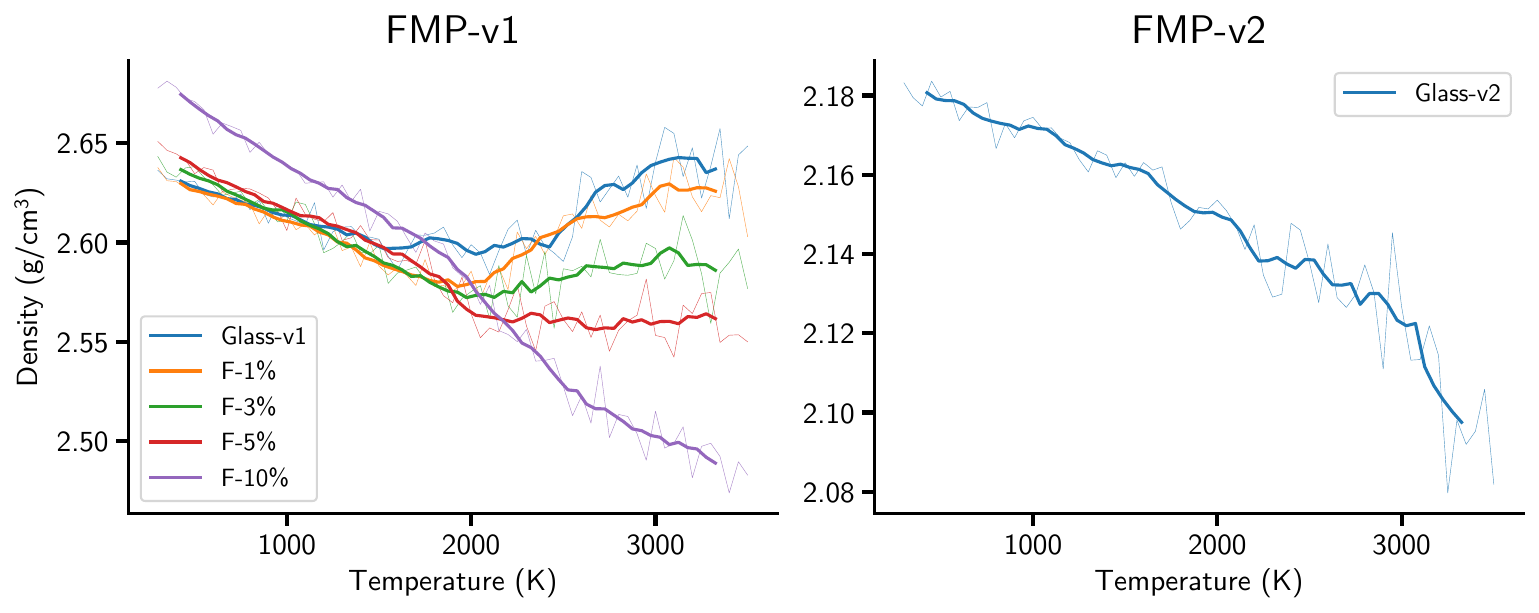}
    \end{center}
    \caption{
    Density as a function of temperature for pure silica and fluorine-doped silica glass systems. Left axis indicates density for systems generated from FMP-v1 (glass-v1 and SiO2-Fx) while right axis indicates density for systems obtained from FMP-v2 (glass-v2). Glass-v1 and glass-v2 both correspond to pure amorphous silica glass systems (\silica{}).
    }
    \label{fig:density}
\end{figure}

Figure \ref{fig:density} shows densities of pure silica glass (\silica) simulated using FMP-v1 and FMP-v2. These \silica{} systems simulated with FMP-v1 and FMP-v2 will be referred to as glass-v1 and glass-v2, respectively. Silica glass systems doped with different concentrations of fluorine atoms are also shown. They are denoted as \silica-Fx, where x represents the relative concentration of oxygen atoms replaced by fluorine. All \silica-Fx systems are simulated only using FMP-v1. Fig \ref{fig:density} shows that glass-v1 exhibits the density anomaly, in which the density decreases with temperature but reaches a minimum around 2150 K, followed by an increase in density until it reaches a plateau around 3500 K. In general, densities of materials monotonically decrease with increasing temperature due to positive thermal expansion coefficients. However, some systems such as silica and water display anomalous density behavior where the thermal expansion coefficients are negative at some range of temperatures higher than their melting points. Contrary to glass-v1, glass-v2 shows a monotonic decrease in density with increasing temperature even though the two potentials have been optimized based on similar DFT data \cite{urata2021a, urata2021b}. 

The simulations are also able to reproduce interesting anomalous density behavior in \silica-Fx systems. With increasing fluorine content, the magnitude in density anomaly decreases. When more than 10 wt\% of oxygen atoms were substituted with fluorine atoms, density anomaly could no longer be observed. In previous studies, it was thought that fluorine doping disrupts the silica network \cite{urata2021a} and reduces the unstable ring structures formed by Si-O-Si networks\cite{Shimodaira2002}, which in turn lessens the density fluctuations in silica. To understand the structural disparity related to such a peculiar phenomenon, we contrasted local topological environments within glass-v1, glass-v2, and fluorine-doped systems.

\subsection{SOAP representation of local environments} \label{subsec:soap}

\begin{figure}[th]
    \begin{center}
    \includegraphics[width=\linewidth]{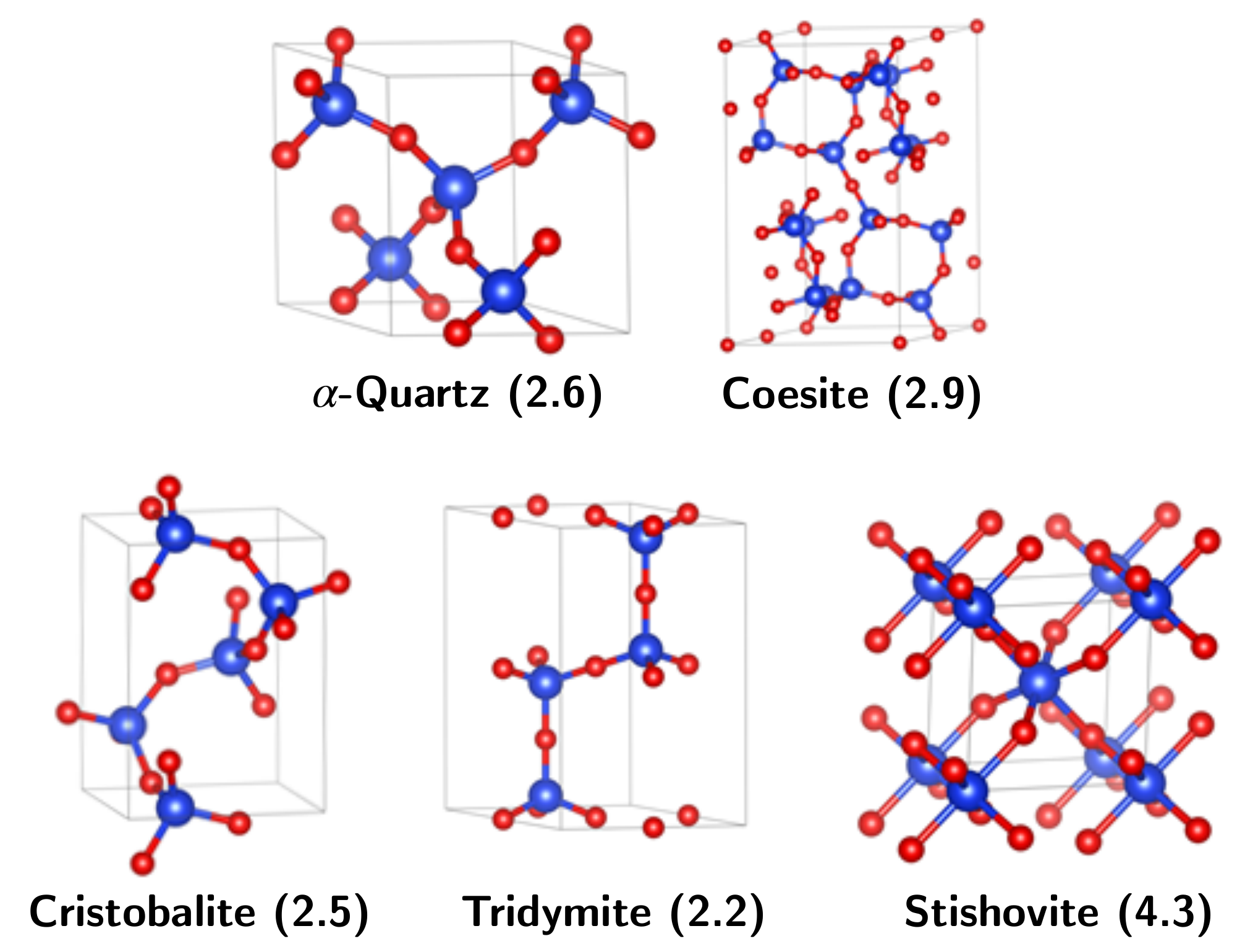}
    \end{center}
    \caption{
    Unit cells of crystalline silica phases used as references in understanding local structural environments of the silica systems. Numbers in the parentheses are approximate densities in $g/cm^3$.
    }
    \label{fig:crystals}
\end{figure}

As a starting point to understand structural differences related to density anomaly behavior, we characterized local environments in glass-v1 and glass-v2 using the SOAP representation \cite{bartok2013}. The SOAP representation takes into account both atomic composition and geometric arrangements, and can be understood in a simpler context as a symmetrized three-body atom correlation function for a local structure centered around an atom\cite{bartok2013,helfrecht_2019}. In our case, we construct SOAP representations that only includes silicon atoms, and any silicon atom outside a cutoff radius of 5~\angstrom{} from a center silicon atom is excluded (for a higher cutoff radius, see Supplementary Information).

While the SOAP representation can encode geometric features of a local environment well, it is best interpreted in combination with dimensionality reduction algorithms. To this end, we applied Uniform Manifold Approximation and Projection for Dimension Reduction (UMAP) to reduce the dimensionality of the representation and to assess the similarity between the environments \cite{mcinnes_2018}. Here, a cosine similarity measure is used. For reference, we also evaluated SOAP representations for crystalline silica phases, namely \quartz{}, tridymite, cristobalite, coesite, and stishovite (Fig. \ref{fig:crystals}). SOAP representations for these phases are constructed by creating supercells, removing all oxygen atoms, choosing a center silicon atoms and removing all silicon atoms outside a cutoff radius of 5~\angstrom{}. A 5~\angstrom{} cutoff radius is chosen since it adequately includes third-order neighbor shells and allows us to get insights into the medium-range order existing in glasses.

\begin{figure*}[t!]
    \centering
    \includegraphics[width=\linewidth]{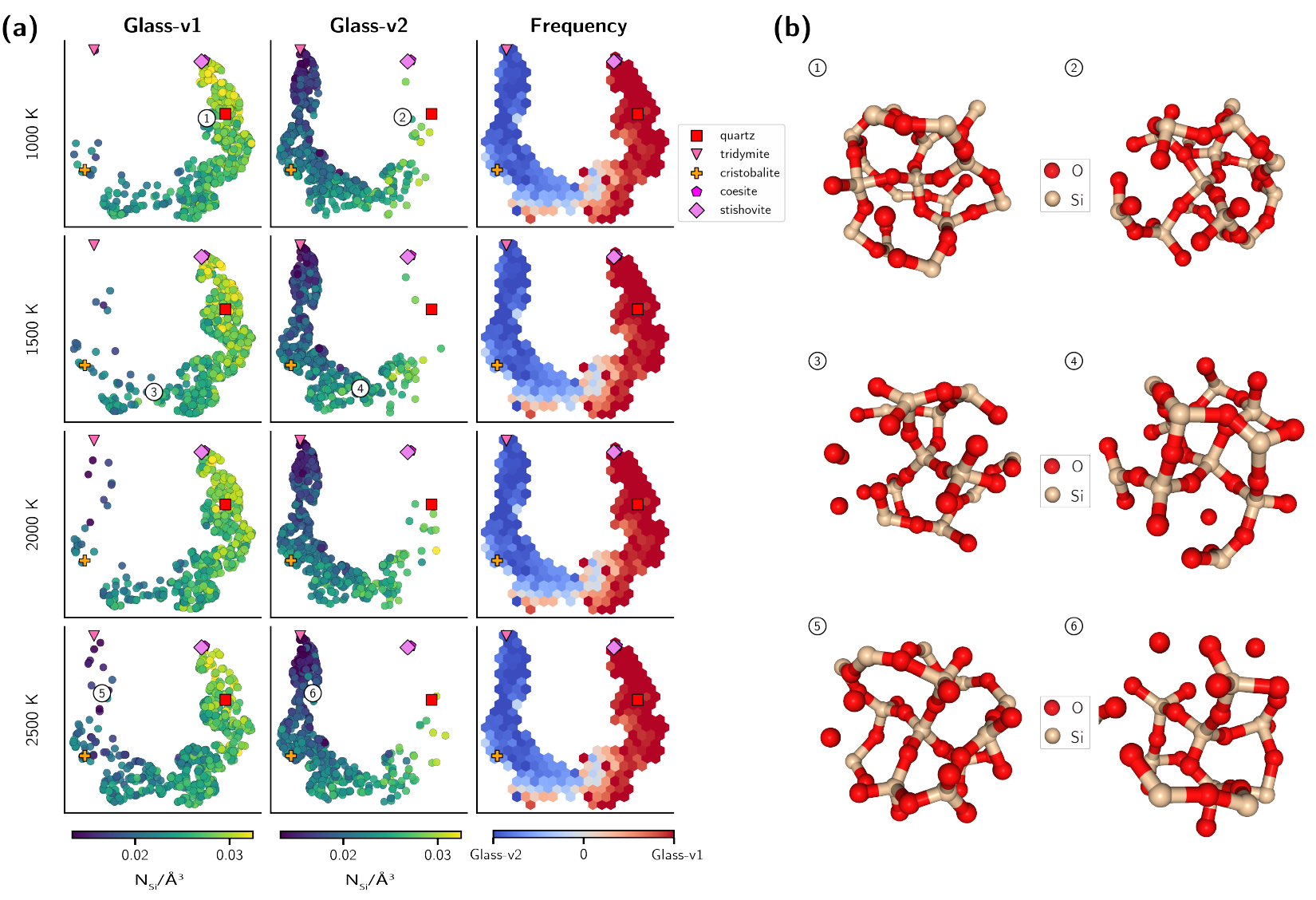}
    \caption{
    \textbf{(a)}, The distribution of local atomic environments in glass-v1 and glass-v2, plotted as points in a two-dimensional space formed by two principal components of UMAP on SOAP representation. The points are colored according to the number density of silicon atoms within a 5~\angstrom{} sphere of the corresponding environment. Each row represents systems at a specific temperature, indicated to the left side. In the third column, a comparison between distributions of the two glass systems is plotted, where the points are colored according to the relative occurrence of atomic environments from either glass-v1 or glass-v2 (i.e. regions with darker red color indicates that more atomic environments from glass-v1 than glass-v2 are distributed there). For reference, SOAP constructions of crystalline phases are highlighted on the plots with specific markers. 
    \textbf{(b)}, Snapshots of local environments in the glass systems from different regions of the SOAP construction plots. Positions of these environments are indicated in (a) with numbers.}
    \label{fig:soap}
\end{figure*}

Figure \ref{fig:soap}a shows the distribution of local environments of glass-v1 and glass-v2 in a two-dimensional space formed by the first two principal components of UMAP reduction based on the SOAP representation. The visualization charts the atomic environments in different regions of the reduced space, indicating that glass-v1 and glass-v2 have dissimilar local environments with only a small overlap. It is important to note that coesite and stishovite are mapped on top of each other in the UMAP projection despite having very different structures, which is possibly due to the lack of high density environments in our systems to effectively distinct these two structures. However, varying UMAP hyperparameters can occasionally change distribution of the projections. glass-v1 has a large amount of local environments similar to that of \quartz{}, which is expected since glassy systems often share similar local structural motifs as their crystalline counterparts despite being disordered. Few environments are also similar to cristobalite, which has a slightly lower density than \quartz{}. There is some number of environments similar to that of coesite and stishovite, but it also shows that not many structures have such high connectedness in the absence of extreme pressures. With increasing temperature, it is observed that the low-density environments akin to that of tridymite increase. Based on the silica phase diagram, increase of tridymite-like local structures is expected. Nevertheless, this is an interesting phenomenon since we expect a decrease in lower density structures at 2500~K due to the anomalous density behavior. 

For glass-v2, the local environment distribution is very different. The majority of environments has low density and hence is more similar to that of cristobalite and tridymite. Interestingly, there are only a few environments that resemble that of \quartz{}. This may be due to the high correlation of SOAP representation to the density, as can be observed from the colors of points in Fig \ref{fig:soap}, even though the interatomic distances of glass-v1 and glass-v2 are similar (see Fig S1 in Supporting Information). When local structures from different regions of the SOAP distribution are visualized (Fig \ref{fig:soap}b), it was found that some structures with similar local motifs can have very dissimilar SOAP representations due to the density difference (Fig \ref{fig:soap}b structures 2 and 5). In this case, despite the ability of SOAP in distinguishing local environments, disentangling density effects from the representation prevents a fair comparison between glass-v1 and glass-v2. This hinders effects in clarifying which local structural features are responsible for the density anomaly of silica glasses, and their correlation with the force-matching potentials.

\subsection{D-measure of local environments} \label{subsec:local_dmeasure}

While SOAP provides structural insight into these amorphous structures, it is very sensitive to density and may show poor sensitivity to connectivity and network effects that are better captured by a discrete connectivity graph. We propose a straightforward, graph-based method to bypass effect of density and to study the (dis)order of glasses beyond the local environment by comparing the network connectivity. Atoms in the local environments are modeled as nodes within undirected graphs, and bonds between two atoms are modeled as the edges. Similar to SOAP construction, local environments are set up as 5~\angstrom{} spheres and any atoms outside of this cutoff radius would not be considered. Hence, each undirected graph represents a local environment, and only silicon atoms are considered as nodes. Two silicons are defined to be connected when they are linked through a bridging oxygen atom. To minimize effects of density differences between glass-v1 and glass-v2, we define a Si-O cutoff distance of 1.9~\angstrom{}. This creates an larger margin for Si-O bond distance, which is defined to be ~1.6~\angstrom{} in crystalline silica materials \cite{Baur1977,Ohsaki1994}.

\begin{figure}[thb]
    \begin{center}
    \includegraphics[width=\linewidth]{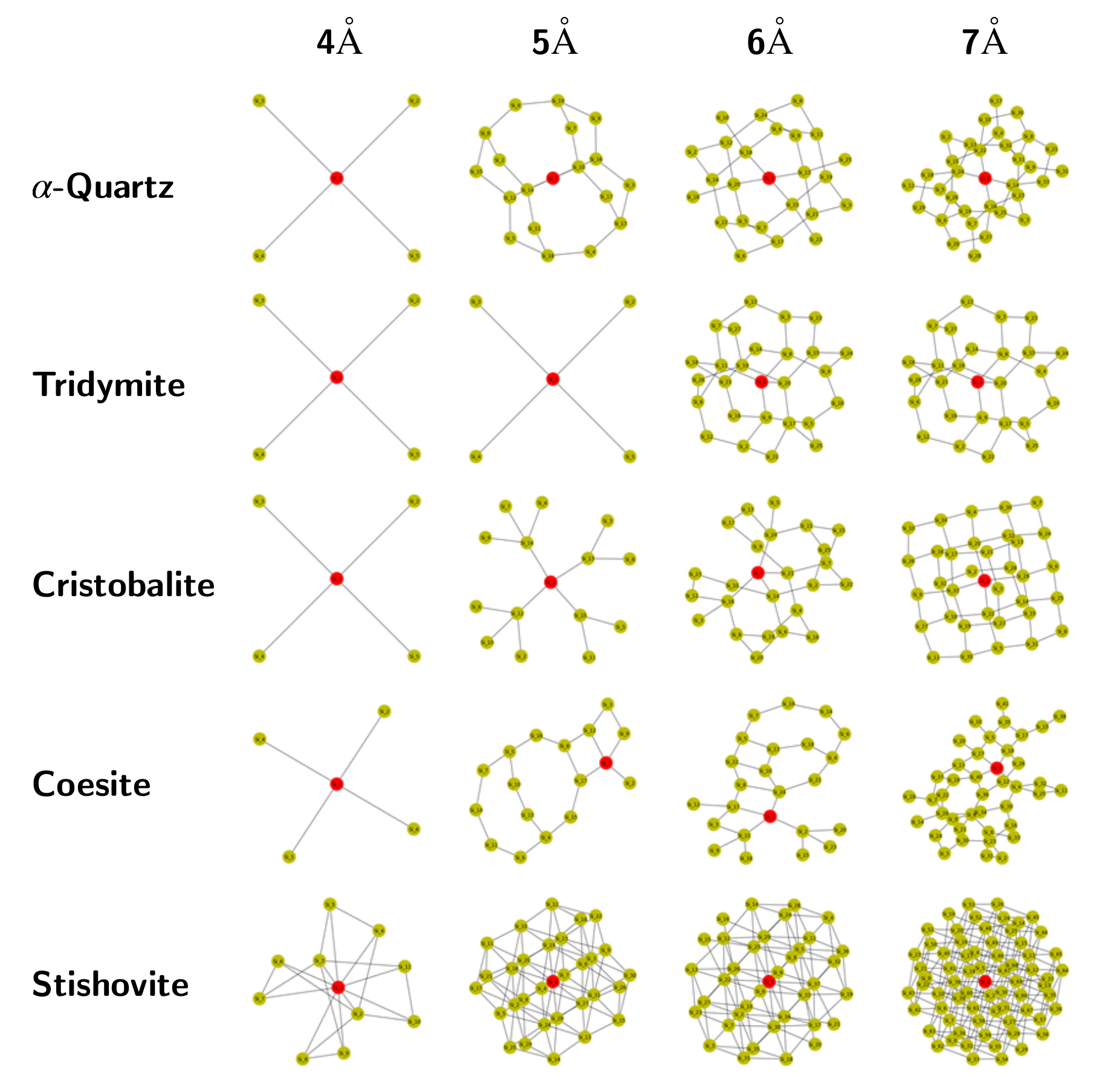}
    \end{center}
    \caption{
    Sub-graphs extracted from the crystalline silica phases (Fig \ref{fig:crystals}) with cutoff distances ranging from 4~\angstrom{} to 7~\angstrom{}. Red spheres are the central silicon atom, while yellow spheres show silicon atoms within the cutoffs. Note that the choice of central atoms does not yield different structures under periodicity conditions for \quartz, tridymite, cristobalite and stishovite. For coesite, two symmetrically-distinct local atomic environments exist\cite{Levien1981} (see Fig S2 in Supporting Information). In our case, only one atomic environment from coesite is used for all subsequent studies. 
    }
    \label{fig:crystal_graphs}
\end{figure}

With the local environments modeled as graphs, we apply a continuous graph dissimilarity metric known as D-measure \cite{schieber2017} (see Section \ref{sec:methods}). Pairwise D-measure values between all local environments extracted from glass-v1 and glass-v2 systems are evaluated to compare all topological differences. For reference, we also modeled the crystalline silica phases discussed in Section \ref{subsec:soap} as graphs (Fig \ref{fig:crystal_graphs}) and evaluated pairwise D-measure of these crystalline phases against the other local environments. Again, UMAP is applied to reduce the dimensionality of the D-measure values. 

\begin{figure*}[th!]
    \centering
    \includegraphics[width=\linewidth]{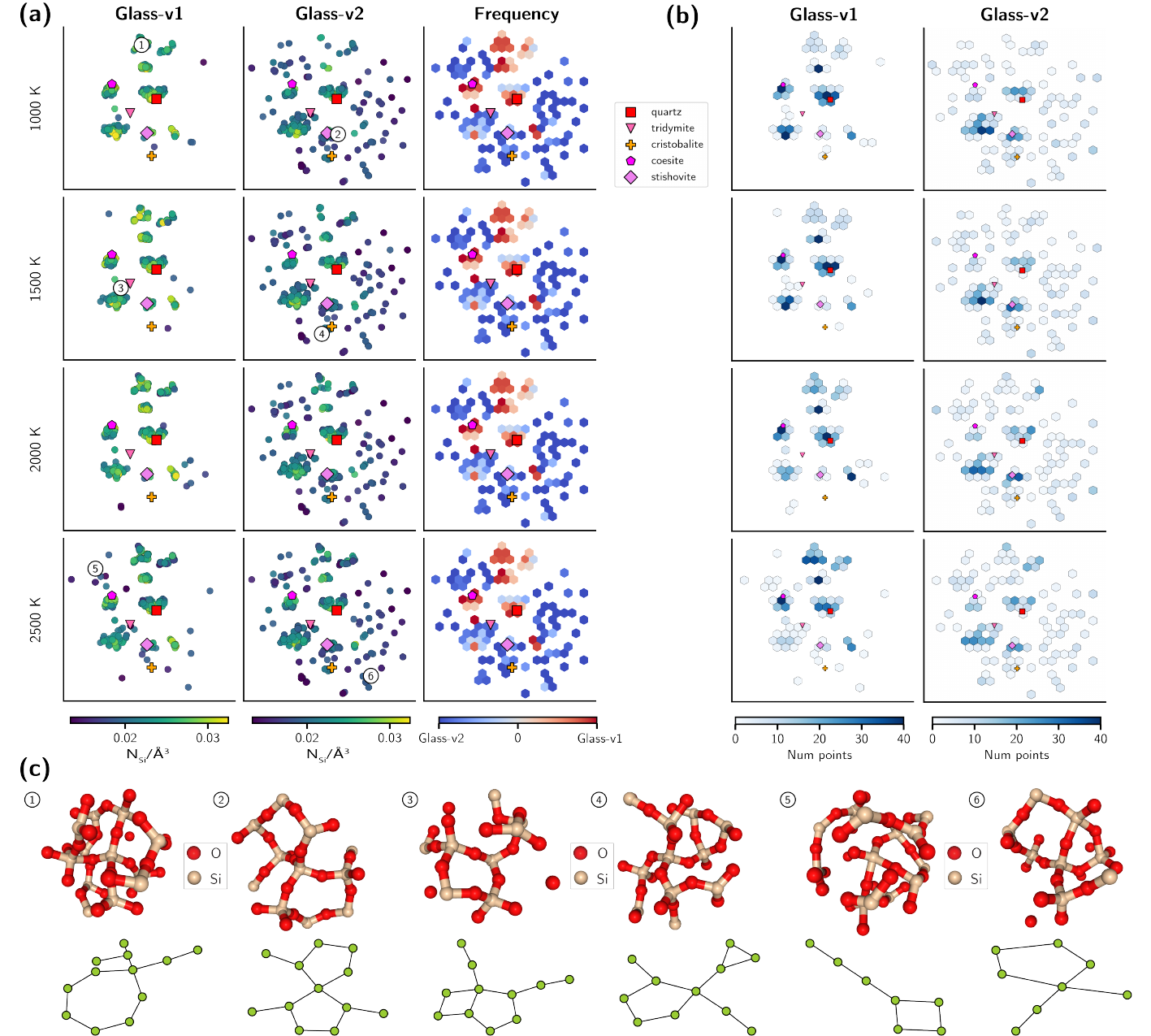}
    \caption{
    \textbf{(a)}, The distribution of local atomic environments in glass-v1 and glass-v2, plotted as points in a two-dimensional space formed by the first two principal components of UMAP on pairwise D-measures. The points are colored according to the number density of silicon atoms within a 5~\angstrom{} sphere of the corresponding environment. Each row represents systems at a specific temperature, indicated to the left side. In the third column, a comparison between distributions of the two glass systems is plotted, where the points are colored according to the relative frequency in occurrence of atomic environments from either glass-v1 or glass-v2 (i.e. regions with darker red color indicates that more atomic environments from glass-v1 than glass-v2 are distributed there). For reference, D-measure constructions of crystalline phases are highlighted on the plots with specific markers. 
    \textbf{(b)}, Distributions of local atomic environments in the first two principal components of UMAP on pairwise D-measure. Axes scales and temperatures of system in each row are the same as (a), but each hexbin is colored according to the number of local environment falling within the area. Note that coordinates of the crystalline phases remain the same but sizes of markers are shrunk for better sight of the hexbins. 
    \textbf{(c)}, Snapshots of local environments in the glass systems from different regions of the D-measure construction plots. Positions of these environments are indicated in (a) with numbers. Constructed graphs are included below the snapshots.
    }
    \label{fig:dmeasure}
\end{figure*}

Fig. \ref{fig:dmeasure}a shows the distribution of local environments of glass-v1 and glass-v2 in a two-dimensional space formed by the first two principal components of UMAP based on the pairwise D-measure values. Unlike for SOAP representation, this results show that there is a substantial overlap in local environments, even if the volumetric densities are different. This illustrates that the graph-based analysis can enable fair comparison between local environments of the systems despite being generated by different force-matching potentials and having different atomic densities. We also recognize the fact that some low density local environments in glass-v2 are mapped as dissimilar environments, which could possibly be caused by the artifact of their sparse density or their unique environment. 

Contrary to the SOAP representation, this graph-based method reveals that glass-v2, to a lesser degree than glass-v1, has an abundance of local environments similar to that of \quartz{} (see Supporting Information for more visualized structures). Intriguingly, densities of points around that of stishovite and coesite crystalline phases are also non-negligible for both glass-v1 and glass-v2, even though none of the local environments have such high densities. Further investigation shows that local structures around stishovite (Fig \ref{fig:dmeasure}c structure 2) and coesite (see Supporting Information) phases generally have a denser number of (inter)connected rings, and increasing temperature does not decrease the amount of such local environments. Conversely, increasing temperature does increase number of local environments in glass-v1 similar to that of lower density phases such as tridymite (Fig \ref{fig:dmeasure}c structure 3) and cristobalite (Supporting Information), which can also be observed in SOAP representation. Other low density structures like Fig \ref{fig:dmeasure}c structure 5 also appear in glass-v1 with increasing temperature, but there is a lack of structures like Fig \ref{fig:dmeasure}c structure 6 perhaps due to asymmetry of structures in a 5~\angstrom{} sphere. 

In order to better understand the density anomaly behavior of silica, the number of points (local environments) falling within specific regions of the UMAP projection are pooled and plotted as colored hexbins, as shown in Fig \ref{fig:dmeasure}b. Via comparison between glass-v1 and glass-v2 environments at 1500 K, 2000 K, and 2500 K, it can be observed that there are distinctly higher numbers of environments similar to structure 1 (Fig \ref{fig:dmeasure}) and the coesite phase in glass-v1, which could be attributed to the anomalous density behavior observed. In general, structures in these two regions have a greater number of (inter)connected rings, but that is not always the case. Some of these structures do not form any rings but have an average connectivity higher than other structures (Fig S6 in Supporting Information). However, it is also worth noting that while we observe such differences between glass-v1 and glass-v2, we cannot be certain of the cause of the density minimum in glass-v1 solely via analysis of plots at temperature 1500~K and 2500~K against 2000 K.

\begin{figure*}[th!]
    \begin{center}
    \includegraphics[width=\linewidth]{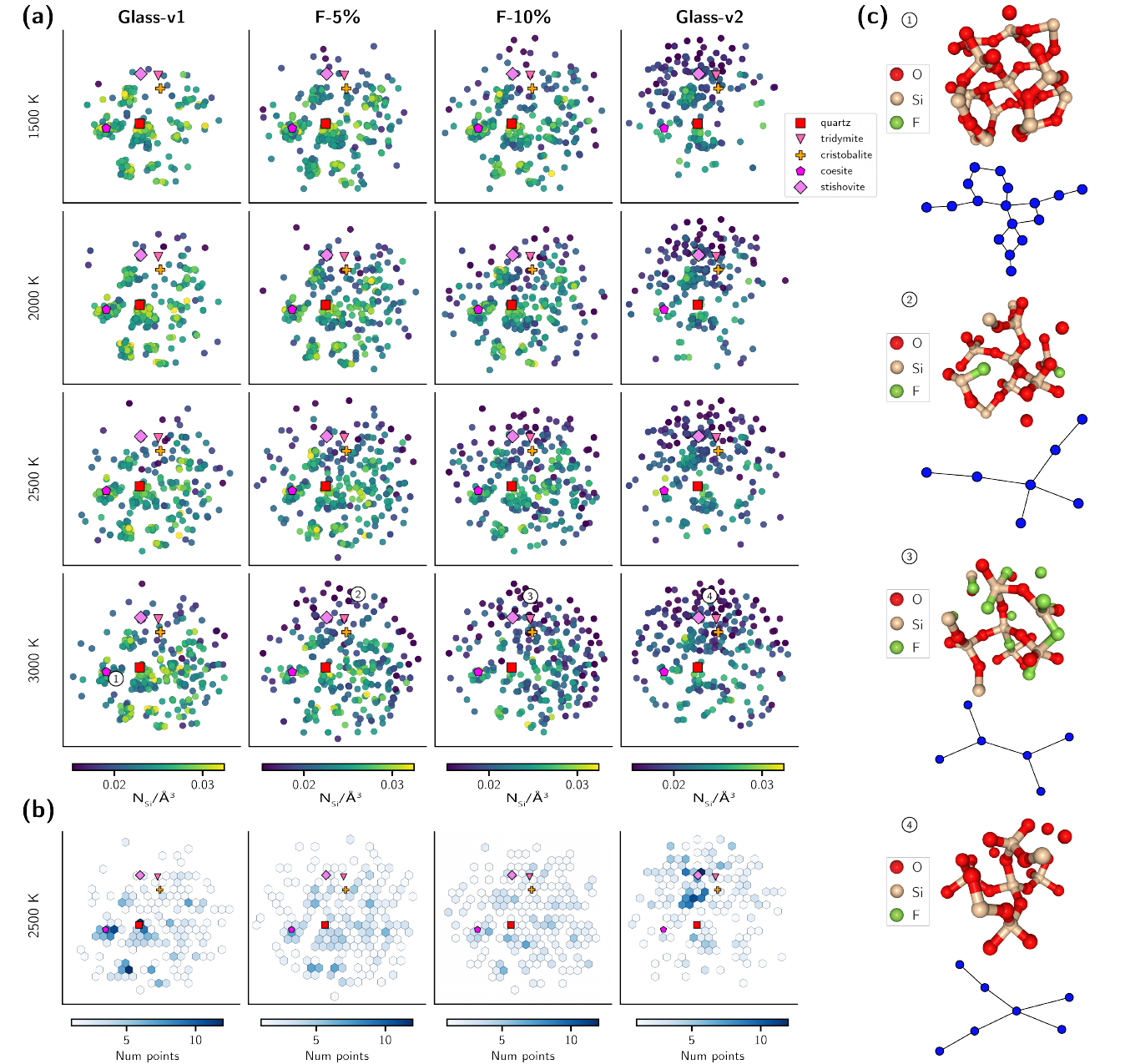}
    \end{center}
    \caption{
    \textbf{(a)}, The distribution of local atomic environments in glass-v1, F-5\% (v1), F-10\% (v1), and glass-v2, plotted as points in a two-dimensional space formed by the first two principal components of UMAP on pairwise D-measure constructions. The points are colored according to the number density of silicon atoms within a 5~\angstrom{} sphere of the corresponding environment. Each row represents systems at a specific temperature, indicated on the left side. For reference, D-measure constructions of crystalline phases are highlighted on the plots with specific markers. 
    \textbf{(b)}, Distribution of local atomic environments in the first two principal components of UMAP on pairwise D-measure. Axes scales are the same as (a), but each hexbin is colored according to the number of local environment falling within the hexbin area. Only systems at 2500~K are shown. Note that coordinates of the crystalline phases remain the same but sizes of markers are shrunk for better sight of the hexbins. 
    \textbf{(c)}, Snapshots of local environments in the glass systems from different regions of the D-measure construction plots (a). Positions of these environments are indicated with numbers. Constructed graphs are included below the snapshots.
    }
    \label{fig:fluorine}
\end{figure*}

To investigate the structural behavior around the density minimum in silica glasses, we examined pairwise D-measure of environments in fluorine-doped (F-doped) systems as increasing F-doping decreases the density anomaly effect, as shown in Fig \ref{fig:density}. Results of pairwise D-measure comparison between glass-v2, F-5\%, F-10\%, and glass-v2 projected using UMAP is thus illustrated in Fig \ref{fig:fluorine}. At 1500 K, with increasing F-doping, local environments with lower silicon density increase, as depicted in top part of the subfigures, even though the overall density of the glass-v1, F-5\% and F-10\% systems are relatively similar (Fig \ref{fig:density}). A possible explanation is that added fluorine atoms terminate connections between silicon atoms by forming Si-F...Si instead of Si-O-Si. Such formation increases the interatomic distances between silicon and oxygen atoms, and with the graph construction method used here captures these terminated connections.

In systems generated by FMP-v1 (glass-v1, F-5\%, and F-10\%) and FMP-v2 (glass-v2), there are noticeable differences in the relative population of stishovite-like, coesite-like, and \quartz{}-like environments (Fig. \ref{fig:fluorine}b). At 2500 K, the number of coesite-like and \quartz{}-like topologies in systems generated by FMP-v1 far exceeds those in glass-v2 systems. Furthermore, as the amount of fluorine-doping increases, quantities of those two environments decrease. On the other hand, the quantity of stishovite-like substructures is much greater in glass-v2 than in FMP-v1 generated systems. Hence, presence or absence of these three dense phases, which are generally characterized by higher number of rings and/or higher connectivity of nodes, could be a starting point for further experimental validation using spectroscopic techniques. Furthermore, it is worth noting that the number of coesite-like structures increases with temperature in glass-v1, which is something not observed in other fluorine-doped or glass-v2 systems, which again could be attributed to the anomalous density behavior.

\subsection{D-measure of global environment} \label{subsec:global_dmeasure}

\begin{figure*}[th!]
    \begin{center}
    \includegraphics[width=\linewidth]{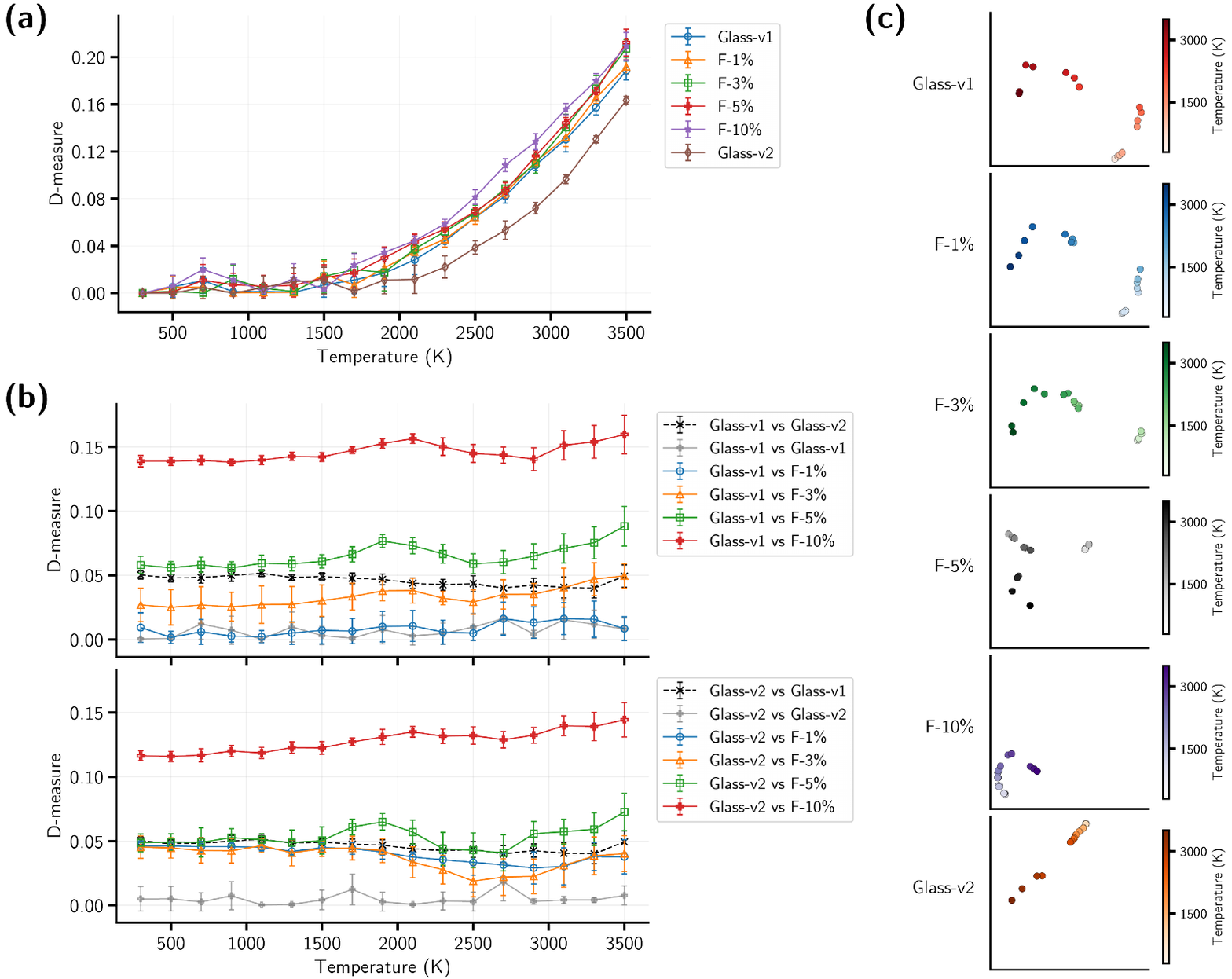}
    \end{center}
    \caption{
    \textbf{(a)}, Change in D-measure values against the initial frames (300 K) of their respective trajectories as a function of temperature for FMP-v1 generated (glass-v1 and fluorine-doped) and FMP-v2 generated (glass-v2) systems.  D-measure values are calculated using graphs constructed from the entire simulation box of the long-time MD trajectories rather than from only a local environment. Connectivity of all silicon atoms within a configuration is taken into account. Points shown are averages of 5 independent MD trajectories with error-bars indicating standard deviation of calculated D-measure values. 
    \textbf{(b)}, Dissimilarity between trajectories of different cases as a function of temperature. Each point shows the average of 25 D-measure values, where frames extracted from 5 independent MD trajectories of one case are compared against frames from 5 trajectories of another case in a pairwise manner. For comparison of the same case (e.g. glass-v1 vs glass-v1), only 10 D-measure values are calculated since similarity between the same frame from identical trajectories would be redundant. Error-bar then indicates standard deviation of the D-measure values. 
    \textbf{(c)}, Distribution of graph configurations for all cases, plotted as points in a two-dimensional space formed by first two principal components of UMAP reduction on pairwise D-measure constructions. Points are colored according to the temperature of the configurations, and only one independent MD trajectory is shown for all cases (see Supporting Information for visualization of all trajectories). 
    }
    \label{fig:global}
\end{figure*}

In addition to pairwise D-measure construction of local environments in the systems, we also compared and analyzed topological changes in the configurations of the entire simulation box using D-measure. SOAP is not computed in this case since difference in densities of the systems would yield very different representations. Again, only silicon atoms are included, and two silicon atoms are considered to be connected when they are bonded by the same oxygen atom. Fig \ref{fig:global}a shows changes in D-measure value of the graph configurations for glass-v1, glass-v2, and different fluorine-doped systems, evaluated against the first frame (300 K) of their respective trajectories. The configurations are taken every 200~K (40~ps) from long-time MD trajectories, and 5 simulations starting from different initial configurations are done for each case. From the figure, we see that D-measure values increase monotonically for all cases, even for FMP-v1 systems that show anomalous density behavior. This implies that the configurations vary continuously with temperature without reversal to topologies that are more similar to their lower-temperature counterparts despite the anomalous density behavior. In addition, the topological transition captured by D-measure for glass-v1 and fluorine-doped systems (FMP-v1 generated) begins to increase at a lower temperature than glass-v2. This is consistent with the fact that fictive temperature of glass-v1, which was determined as a folding point of temperature-potential energy curve, is lower than that of glass-v2 (see Table S1 in the Supporting Information). In this case, it is possible to infer that the structural changes identified by the D-measure may be related to the glass transition of silica glass. Since higher fluorine doping is known to lower the fictive temperature of silica glass further\cite{shiraki1992, kirchhof2018, urata2021a}, D-measure values of the \silica{} systems containing more fluorine also increase at lower temperature. According to these observations, it is well supported that the D-measure appropriately captures the global network changes related to the glass transition in silica glass.

Next, we quantified the variability of the graph configurations between different cases in Fig \ref{fig:global}b. Comparisons between any two cases at a specific temperature consist of 25 D-measure values, where 5 frames from the first case (5 independent MD trajectories) are compared against 5 frames from the second case in a pairwise manner. Based on results discussed in Section \ref{subsec:local_dmeasure}, we would expect glass-v2 to be the most dissimilar to glass-v1 since they show the largest difference in local environment distributions. However, the results demonstrate that higher fluorine doping distorts the overall topology of \silica{} systems more. UMAP of the pairwise D-measure values of the cases shown in Fig \ref{fig:global}c also confirm this finding. With 1 wt\% fluorine doping, distributions of glass-v1 and F-1\% are quite similar. At 3 wt\% fluorine content, distribution starts to deviate from glass-v1, whereas at 10 wt\%, the system has a completely distinct distribution. These findings confirm that fluorine doping changes the overall structure of \silica{}. Glass-v2, on the other hand, is consistently dissimilar from all other systems due to the different potential used, but still has higher similarity in distribution to lower fluorine-doped systems than F-10\% system nonetheless.

\section{Conclusions} \label{sec:conclusions}

In summary, we proposed a novel approach to study the topological changes of network glasses by casting the disordered structures as graphs and applying D-measure to quantify similarities between the formulated graphs. This method can be applied to understand configurations within environments of any range. In this paper, we apply this method to study the structural origin of the density anomaly phenomenon in pure and fluorine-doped amorphous silica (\silica) systems. Atomistic structures from MD simulations generated from two different potentials are compared, only one of which generates silica glass systems that exhibit anomalous density behavior. 

We have found that structural analysis using D-measure appropriately distinguishes similar glass structures encountered during heating progression of the systems and identifies structural differences of the \silica{} systems modeled from different potentials. In contrast to descriptors like SOAP, which vary with density, this approach is able to only focus on topological effects within disordered structures. In applications of this method on short-range local structures, we are able to identify the increase in local orderings similar to phases like coesite and \quartz{} in \silica{} systems that exhibit anomalous density behavior at temperatures around the density minimum ($\sim$2000 K). Using the global topological comparison, however, the results do not show that structures of the systems exhibiting anomalous density behavior reverts to a lower-temperature topology, and vary continuously regardless of the density minima. In addition, we are able to verify terminated connections between Si-O-Si bonds in fluorine-doped \silica{} systems with this method. This is also shown in the global environment comparison between these systems, where a higher fluorine-doping distorts the structure more. All things considered, this study demonstrated the applicability of graph theory to find both short- and long-range structural signatures in an amorphous material and can help with better understanding of elusive properties within disordered systems.

\section{Methods} \label{sec:methods}
\subsection{D-measure} \label{subsec:d-measure}

D-measure, $D(G, G')$ evaluates the dissimilarity between two undirected graphs, $G$ and $G'$. First, the probability distribution function (PDF) of node-distances for every node is generated and then PDFs are compared through the following definitions \cite{schieber2017}.

\begin{eqnarray}
    \label{eq:dmeasure}
    &D&(G,G') = w_1 \sqrt{{J(\mu_G,\mu_{G'}) \over log2}} \\
    &+& w_2 \left| \sqrt{\textrm{NND(G)}} - \sqrt{\textrm{NND(G')}} \right| \\ 
    &+& {w_3 \over 2} \biggl(\sqrt{{J(P_{\alpha G}, P_{\alpha G'}) \over \log 2}}  + \sqrt{{J(P_{\alpha G^C}, P_{\alpha G^{C'}}) \over \log 2}} \biggr),
\end{eqnarray}

where 
\begin{eqnarray}
    \label{eq:jensen-shannon}
    J({\bf P}_1, ... ,{\bf P}_N) &=& {1 \over N} \sum_{i, j} p_i(j) \log \biggl( {p_i(j) \over \mu_j } \biggr)\\
    \label{eq:mu}
    \mu_j &=& {1 \over N} \sum_{i=1}^N p_i(j)\\
    \label{eq:nnd}
    \textrm{NND}(G) &=& { J({\bf P}_1, ... ,{\bf P}_N) \over \log(d+1) }
\end{eqnarray}

$G^C$ and $G^{C'}$ indicate the complement of $G$ and $G'$. The distance distribution is defined as ${\bf P}_i = \{p_i(j)\}$, where ${p_i(j)}$ 
is the fraction of nodes connected to node $i$ at distance $j$. Accordingly, a graph composed of $N$ nodes is written as \{${\bf P}_1$, ... ,${\bf P}_N$\}.     
The first term of Eq. (\ref{eq:dmeasure}) compares graph distributions representing averaged node's connectivity patterns using the Jensen-Shannon divergence (Eq. \ref{eq:jensen-shannon}). The second term evaluates heterogeneity of the nodes in terms of their connectivity profiles via the network node dispersion term, $\textrm{NND}$ (Eq. \ref{eq:nnd}). The third term measures centrality of each node by considering the span of nodes both directly and indirectly connected. 

$D(G, G')$ is constrained within $[0, 1)$ under the limitation that $w_1 + w_2 + w_3 = 1$. Two isomorphic networks return $D(G, G') = 0$, but two non-isomorphic networks will not necessary return non-zero D-measure. Furthermore, graph sizes of $G$ and $G'$ need not be the same. 

In this work, the weights are set to be $w_1 = w_2 = 0.45$ and $w_3 = 0.1$ as the original study suggested \cite{schieber2017}. To simplify the graphs, only silicon atoms are considered as nodes. Two silicons are defined to be connected when they are linked through a bridging oxygen atom, with a Si-O cutoff distance of 1.9~\angstrom{}. 

\subsection{Molecular Dynamics simulations} \label{subsec:md}

Table 1 summarizes all the models constructed by MD simulations. The LAMMPS package \cite{plimpton1995} was employed for the MD simulations. All simulations were conducted in an isothermal-isobaric ensemble (NPT) using Nos\'{e}--Hoover thermostat \cite{nose1984} and barostat \cite{tuckerman2006}. The F-doped silica glass models were constructed by replacing 1, 3, 5 and 10 \% of oxygen of the silica glass model with fluorine. They are denoted as SiO$_2$-Fx (x = 1, 3, 5, 10). The glass models were obtained using the melt-quenching method \cite{urata2017} where initial glass structures were melted at 3500~K for 2.5~ns, and 5 different configurations were extracted every 500~ps to act as the starting configurations for 5 different MD trajectories. Each of the 5 configurations were cooled down from 3500~K to 300~K at a cooling rate of 1~K~ps$^{-1}$, followed by equilibration at 300~K for 500~ps each. They were then heated up again to 3500~K at a rate of 0.5~K~ps$^{-1}$. Individual glass models were recorded every 5~K (10~ps) and each model was then equilibrated for 10~ps at the corresponding temperatures. Only data from the last 3~ps was used to average the properties, such as density and potential energy. These five independent replicas were examined for all the glass models to verify reproducibility. In all cases, equations of motion were integrated with a time step of 1.0 fs.

Two force-matching potentials (FMP) \cite{urata2021a, urata2021b} were utilized. These have been reported elsewhere to reproduce forces from density functional theory (DFT) calculations performed on \silica{} and F-doped \silica. The configurations used for DFT calculations were extracted from trajectories of long-time classical MD simulations performed
using the Buckingham-type functional in the Teter potential \cite{du2004}. The first FMP (named as FMP-v1 hereafter) reproduces density minima of silica at a lower temperature than the BKS potential, but overestimates density of silica at room temperature\cite{urata2021a}. The second FMP (FMP-v2) was developed by considering both forces of atoms and energies of configurations to remedy the overestimation of density in FMP-v1 \cite{urata2021b}. Interestingly, FMP-v2 is able reproduce the room temperature density of silica glass accurately, but does not exhibit density anomaly of silica glass.

\begin{table}[thb]
\caption{Glass models for molecular dynamics simulations. 
}
\centering
\small
\begin{tabular}{ l c c c c c c }
\hline\noalign{\smallskip}
  Abbreviation & \multicolumn{4}{c}{No of atoms}  & F content &  F/O \\
                     & Si & O & F & Total             & [wt\%] & ratio [\%] \\
\noalign{\smallskip}\hline\noalign{\smallskip}
\silica  & 3333  & 6666  & -  & 9999  & - & - \\
SiO$_2$-F1  & 3333  & 6600  & 132  & 10065 & 1.2 & 2.0 \\  
SiO$_2$-F3  & 3333  & 6467  & 398 & 10298  & 3.7 & 6.2 \\  
SiO$_2$-F5  & 3333  & 6333  & 666 & 10332  & 6.1 & 10.5 \\  
SiO$_2$-F10  & 3333  & 6000  & 1332 & 10665  & 11.8 & 22.2 \\  
\noalign{\smallskip}\hline\noalign{\smallskip}
\end{tabular}
\label{tab1}   
\end{table} 

\section{Data Availability}
The molecular dynamic trajectories generated in the current study have been deposited in the Materials Cloud Archive under accession code \url{https://doi.org/10.24435/materialscloud:yv-e7}.\cite{tan2021data}

\section{Code Availability}
The code used to produce the results in this paper is from a previous paper: Graph similarity drives zeolite diffusionless transformations and intergrowth where the code is available at \url{https://github.com/learningmatter-mit/Zeolite-Graph-Similarity/blob/master/zeograph/dmeasure.py} under the MIT license. \cite{schwalbe2019}

\section{Acknowledgements}
This work was supported by funding from the AGC, Inc.

\section{Competing Interests}
The authors declare no competing interests.

\section{Author Contributions}
A.R.T. and S.U. contributed equally. A.R.T. and S.U. designed the experiments, performed data analysis and wrote the manuscript. M.Y. assisted with code implementation for D-measure. R.G.-B. supervised the research, assisted in results interpretation and contributed to manuscript writing. 

\section{Materials \& Correspondence}
Correspondence to Rafael G\'omez-Bombarelli and Shingo Urata.

\newpage
\section*{References}

\bibliography{references}

\foreach \x in {1,...,14}
{%
\clearpage
\includepdf[pages={\x}]{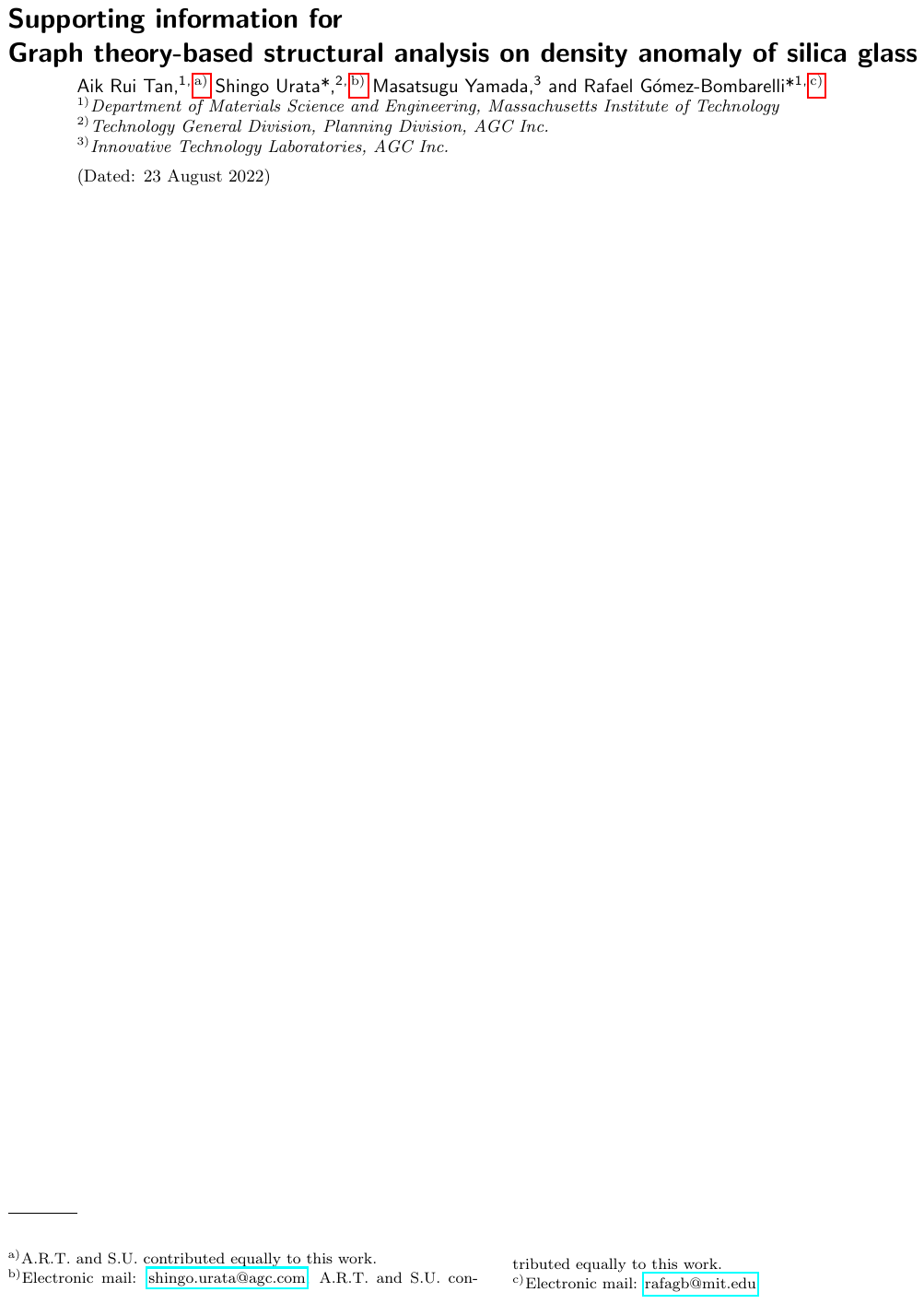} 
}

\end{document}